\documentclass[journal,twoside,web]{ieeecolor}

\usepackage{generic}
\usepackage{cite}
\usepackage{amsmath,amssymb,amsfonts}
\usepackage{algorithmic}
\usepackage{graphicx}
\usepackage{textcomp}
\usepackage{hyperref}
\usepackage{graphicx}
\usepackage{subfigure}
\usepackage{epsfig} 
\usepackage{cite}
\usepackage{lcsys}

\newtheorem{theorem}{Theorem}[section]
\newtheorem{lemma}[theorem]{Lemma}
\newtheorem{proposition}[theorem]{Proposition}

\newtheorem{definition}[theorem]{Definition}
\newtheorem{remark}[theorem]{Remark}
\newtheorem{assumption}[theorem]{Assumption}

\begin{document}

\def\BibTeX{{\rm B\kern-.05em{\sc i\kern-.025em b}\kern-.08em
    T\kern-.1667em\lower.7ex\hbox{E}\kern-.125emX}}
\markboth{\journalname, VOL. XX, NO. XX, XXXX 2017}
{Author \MakeLowercase{\textit{et al.}}: Preparation of Papers for IEEE Control Systems Letters (August 2022)}

\title{Data-Driven Input-Output Control Barrier Functions}

\author{Mohammad Bajelani and Klaske van Heusden
\thanks{We acknowledge the support of the Natural Sciences and Engineering Research Council of Canada (NSERC) [RGPIN-2023-03660].}
\thanks{The University of British Columbia, School of Engineering, 3333 University Way, Kelowna, BC V1V 1V7, Canada {{\tt\small mohammad.bajelani, klaske.vanheusden @ubc.ca}}}}

\maketitle
\thispagestyle{empty}


\begin{abstract}
    Control Barrier Functions (CBFs) offer a framework for ensuring set invariance and designing constrained control laws. However, crafting a valid CBF relies on system-specific assumptions and the availability of an accurate system model, underscoring the need for systematic data-driven synthesis methods. This paper introduces a data-driven approach to synthesizing a CBF for discrete-time LTI systems using only input-output measurements. The method begins by computing the maximal control invariant set using an input-output data-driven representation, eliminating the need for precise knowledge of the system's order and explicit state estimation. The proposed CBF is then systematically derived from this set, which can accommodate multiple input-output constraints. Furthermore, the proposed CBF is leveraged to develop a minimally invasive safety filter that ensures recursive feasibility with an adaptive decay rate. To improve clarity, we assume a noise-free dataset, though the method can be extended using data-driven reachability to capture noise effects and handle uncertainty. Finally, the effectiveness of the proposed method is demonstrated on an unknown time-delay system.
\end{abstract}

\section{Introduction}

\IEEEPARstart{S}{afety-critical} applications require constrained control laws to guarantee constraint satisfaction. Control Barrier Functions (CBFs) are widely employed to design such control laws and ensure set invariance within the state-space framework. In the continuous-time setting, a CBF is generally defined as a Lyapunov-like function that remains positive on its zero-level set and guarantees invariance by maintaining a non-negative rate at the boundary of this set. However, identifying a \emph{valid} CBF can be challenging and case-specific, particularly when addressing input constraints, high relative orders, or multiple state constraints. Extensions to address these scenarios are discussed in \cite{breeden2023compositions, agrawal2021safe, nguyen2016exponential}, while comprehensive overviews of CBFs are provided in \cite{ames2016control, cohen2024safety_review}. This emphasizes the need for a \emph{systematic method} to design CBFs, a topic that has gained significant attention recently. We classify these methods into four broad categories: CBFs using function approximation techniques, set-driven CBFs, backup CBFs, and CBFs based on reduced-order models. Each approach is based on different assumptions and exhibits varying levels of conservatism in ensuring safety, along with differences in offline and online computational load.

Function approximation techniques assume that a valid CBF exists within a specific function space and aim to identify it. For instance, sum-of-squares programming represents CBFs as polynomial terms, addressing the barrier conditions via semi-definite programming \cite{clark2021verification, prajna2007framework}. Another approach utilizes neural networks, which can provide less conservative solutions but require large datasets \cite{wang2024simultaneous}. Additionally, function approximators such as Koopman models and Gaussian process models have been applied in CBF construction \cite{wang2024simultaneous, yaghoubi2020training, 10156243, jagtap2020control}.

Alternatively, set-driven CBFs assume that a control-invariant set is computed using reachability analysis techniques, which allows for the direct construction of the CBFs. For instance, a CBF is proposed by computing the zero-level set of the value function through the solution of the Hamilton-Jacobi-Bellman (HJB) equation in \cite{choi2021robust}. In contrast, a method proposed in \cite{freire2023systematic} directly constructs a discrete-time CBF using maximal output admissible sets, often leading to more conservative solutions.

For high-dimensional nonlinear systems with layered control architectures, designing CBFs can be challenging. One effective approach is constructing CBFs for reduced-order models and extending the results to higher-dimensional systems through a tracking controller \cite{cohen2024safety, molnar2023safety}. Like Model Predictive Safety Filters (MPSFs), backup CBFs start with a conservative CBF and use system dynamics to compute a backup trajectory by solving an initial value problem  \cite{folkestad2020data}. 

Despite significant progress in systematically designing CBFs, no existing method directly constructs them from input-output data. Most approaches rely on highly accurate models, extensive experimentation, and full-state measurements, often assuming minimal measurement noise. For real-world applications, these methods may need to be extended to the input-output framework, a task that can complicate the design due to state estimation errors.  Moreover, some methods derive a candidate CBF while neglecting input constraints, which can jeopardize recursive feasibility and lead to violations of input-output constraints in practical applications.

This paper introduces a systematic method for crafting Input-Output Discrete-time Control Barrier Functions (IO-DCBFs) for LTI systems using a single input-output trajectory dataset. This method handles state estimation by treating past measurements as an extended state, utilizing the maximal control-invariant set and a data-driven representation to ensure minimally conservative and recursively feasible solutions. Notably, time delays are managed through the input-output data-driven approach, and the exact system order is not required. Additionally, we show that the proposed CBF is equivalent to MPSF under specific configurations. The methodology is established in a deterministic setting, which can be viewed as the data-driven counterpart of model equivalence in model-based safety filters. 

The paper is structured as follows: Section \ref{sec: Preliminaries and Problem Statement} introduces the input-output framework for constrained control, outlining the preliminary material, key assumptions, and a data-driven representation. Section \ref{sec: Method} details the main contribution, proposing a CBF from input-output data. In Section \ref{sec: VS}, we define the terminal components and prediction horizon of MPSF, as well as the decay rate for the proposed control barrier function, highlighting their equivalence. Section \ref{sec: sim} provides a numerical example to showcase the effectiveness of the proposed method. Finally, Section \ref{sec: discussion} presents the discussion and conclusion.
\section{A Data-Driven Input-output Representation for constrained LTI systems} \label{sec: Preliminaries and Problem Statement}
Assume the true system is a deterministic, discrete-time LTI system, described in minimal state-space form:
\begin{subequations} \label{eq: LTI-system}
    \begin{equation}
        x_{t+1} = A x_t + B u_t, \quad y_t = C x_t,
    \end{equation}
    \begin{equation}  \label{eq:polytopic constraints}
        u_t \in \mathcal{U}, \quad  y_t \in \mathcal{Y},
    \end{equation}
\end{subequations}
where $x_t \in \mathbb{R}^n$, $u_t \in \mathbb{R}^m$, and $y_t \in \mathbb{R}^p$ are the state, input, and output vectors at time step $t$, respectively. The sets $\mathcal{U}$ and $\mathcal{Y}$ are polytopes that define the admissible input and output constraints. Let the pair \((A,B)\) be controllable and \((A,C)\) observable, and the tuple \((A,B,C)\) be unknown in both its values and dimensions. Assume that states cannot be measured and an input-output trajectory dataset of the system \eqref{eq: LTI-system} with length \(N_0\) is available in the form of the following vectors:
\begin{subequations} \label{eq: IO dataset}
    \begin{equation}
    u_{\left[0, N_0-1\right]}^d = \left[u_0^{\top}, \ldots, u_{N_0-1}^{\top}\right]^{\top} \in \mathbb{R}^{mN_0\times1},
    \end{equation}
    \begin{equation}
    y_{\left[0, N_0-1\right]}^d = \left[y_0^{\top}, \ldots, y_{N_0-1}^{\top}\right]^{\top} \in  \mathbb{R}^{pN_0\times1}.
    \end{equation}
\end{subequations}

The Hankel matrices \(H_L(u^d) \in \mathbb{R}^{(mL) \times (N_0 - pL + p)}\) and \(H_L(y^d) \in \mathbb{R}^{(pL) \times (N_0 - pL + p)}\), corresponding to the given input-output trajectory and consisting of \(mL\) and \(pL\) rows, are defined as follows:
\begin{subequations} \label{eq: Hankel matrices}
    \begin{equation}
    H_L(u^d) = \big[ u^d_{[0:L-1]}, u^d_{[1:L]}, \dots, u^d_{[N_0-L:N_0-1]} \big],
    \end{equation}
    \begin{equation}
    H_L(y^d) = \big[ y^d_{[0:L-1]}, y^d_{[1:L]}, \dots, y^d_{[N_0-L:N_0-1]} \big],
    \end{equation}
\end{subequations}
where, each column \(u_{[t:t+L-1]}\) and \(y_{[t:t+L-1]}\) represents a sliding window of \(L\) consecutive stacked inputs or outputs, starting at time \(t\).
\begin{definition}[System's Lag \cite{9654975}] \label{definition: lag} 
    $\underline{l}=l(A, C)$ denotes the lag of the system (\ref{eq: LTI-system}), in which $l(A, C)$ is the smallest integer that can make the observability matrix full rank.
    \begin{equation*}
        l(A, C):=(C, C A, \ldots, C A^{l-1}).
    \end{equation*}
\end{definition}
\begin{definition}[Persistently Excitation] \label{Def: PE condition}
    Let the Hankel matrix's rank be $rank(H_L(u^d)) = mL$, then $u^d$ represents a persistently exciting signal of order $L$.
\end{definition}

The behavioral system theory allows us to represent dynamical systems using time-series data, meaning the representation is purely data-driven and not confined to a specific model. For LTI systems, it establishes identifiability conditions, demonstrating how the single input-output trajectory dataset \eqref{eq: IO dataset} can exactly describe system \eqref{eq: LTI-system} without under-modeling. Required prior knowledge of the underlying system is an arbitrarily large upper bound on the system's lag, while the dataset in equation \eqref{eq: IO dataset} is assumed to be noise-free and to satisfy the persistent excitation (PE) conditions.
\begin{assumption}[Upper Bound on System's Lag] \label{Assumption: I} An upper bound on the system's lag is known $T_{\text{ini}}\geq \underline{l}$.
\end{assumption}
\begin{assumption}[Persistent Excitation \cite{berberich2020robust}] \label{Assumption: II}
The stacked Hankel matrix, \( [H_u^{\top} H_y^{\top}]^{\top} \), is PE of order $L = T_{\text{ini}} +1$ in the sense of Definition \ref{Def: PE condition}.
\end{assumption}
\begin{lemma}[Fundamental Lemma \cite{willems2005note}] \label{Fundamental Lemma}
    Let Assumptions \ref{Assumption: I}-\ref{Assumption: II} hold. Then, \( [H_u^{\top} H_y^{\top}]^{\top} \) spans all the input-output trajectories of the system \eqref{eq: LTI-system} with the length of \( L \).
\end{lemma}

Let us divide the Hankel matrices \eqref{eq: Hankel matrices} into two parts, named past and future data. In detail, let the first $T_{\text{ini}}$ rows of $H_L(u)$ and $H_L(y)$ be $U_p$ and $Y_p$, and the last row be $U_f$ and $Y_f$, respectively. For any choice of \(T_{\text{ini}} \geq \underline{l}\), the initial condition and the system's order are implicitly determined. In other words, considering a sufficiently long segment of the past input-output trajectory can express the history of the dynamics of system \eqref{eq: LTI-system}, thereby eliminating the need for the underlying state, as noted in \cite[lemma 1]{markovsky2008data}. The past input-output trajectory can be considered as an extended state as suggested in \cite{berberich2021design}. Since the system states in \eqref{eq: LTI-system} are neither accessible nor explicitly constrained, we define the extended input-output constraints as follows:
\begin{definition}[Extended State and Extended Constraints] \label{definition: Extended State}
    For some integers $T_{\text{ini}}\geq\underline{l}~$, the extended state $\xi$ at time $t$ is defined as follows
    \begin{align}\label{eq:extended_state}
    \xi_t := \begin{bmatrix}u^{\top}_{[t-T_{\text{ini}},t-1]}, y^{\top}_{[t-T_{\text{ini}},t-1]}\end{bmatrix}^{\top} \in \mathbb{R}^{(m+p)T_{\text{ini}} \times 1},
    \end{align}
    where $u_{[t-T_{\text{ini}},t-1]}$ and $y_{[t-T_{\text{ini}},t-1]}$ denote the last $T_{\text{ini}}$ input and output measurements at time $t$. Consequently extended state constraint is defined as follows for all $i=\{1,\hdots, T_{\text{ini}}\}$:
    \begin{equation} \label{eq: extended-constraints}
    \Xi=\{\xi_t \in \mathbb{R}^{(m+p)T_{\text{ini}} \times 1}| u_{t-i} \in U , y_{t-i} \in Y\}.
    \end{equation}
\end{definition}

The extended constraint in \eqref{eq: extended-constraints} implies that not only \( u_t \in \mathcal{U} \) and \( y_t \in \mathcal{Y} \), but also all past input-output trajectories of length \( T_{\text{ini}} \) must lie within \( (\mathcal{U}, \mathcal{Y}) \). This leads to the concept of the input-output safe control invariance set, as defined below.
\begin{definition}[Input-output safe control invariant set] \label{def: Input-Output safe Control Invariant Set}
    A set \( \mathcal{S}_\Xi \subseteq \Xi \) is defined as an input-output safe control invariant set for the system (\ref{eq: LTI-system}), if $\forall \xi_{t} \in \mathcal{S}_\Xi$ such that $\exists u_{t} \in \mathcal{U} \text{ and } \xi_{t+1} \in \mathcal{S}_\Xi$.
\end{definition}

To calculate the input-output safe set $\mathcal{S}_\Xi$, we proposed a sample-based data-driven method in \cite{pmlr-v242-bajelani24a} that does not require an explicit data-driven representation. This section employs an explicit data-driven representation to compute the input-output safe set, resulting in a more time-efficient solution. As shown in \cite{qin2024data}, if the system \eqref{eq: LTI-system} is strictly proper, then we can establish a one-step ahead data-driven predictor as follows, where $y_{t}$ can be unambiguously determined.
\begin{equation}
y_{t}= R \xi_t, \quad \forall R \in \mathcal{R},
\end{equation}
where, \(\mathcal{R} := \{R \in \mathbb{R}^{p\times(m+p)T_{\text{ini}}}: R W = Y_f \}\) and \( W = [U_p^{\top} \; Y_p^{\top}] \). Note that \(\xi_t\) lies in the span of \(W\) since it represents an input-output trajectory of system \eqref{eq: LTI-system}. If the system's lag is known (\(T_{\text{ini}} = \underline{l}\)), \(R\) can be uniquely determined; otherwise, for any choice of (\(T_{\text{ini}} > \underline{l}\)), $R$ should only satisfy $R W = Y_f$. The data-driven representation is then defined as follows:
\begin{subequations} \label{eq: one-step data-driven predictor}
\begin{equation}
\xi_{t+1} = A_e \xi_{t} + B_e u_{t},
\end{equation}
\begin{equation}
\setlength{\arraycolsep}{1pt}
A_e = \left[\begin{array}{ccc}
\mathbf{0}_{m(T_{\text{ini}}-1) \times m} & \mathbf{I}_{m(T_{\text{ini}}-1)} & \mathbf{0}_{m(T_{\text{ini}}-1) \times pT_{\text{ini}}} \\
\multicolumn{3}{c}{\mathbf{0}_{m \times (m+p)l}} 
\\
\mathbf{0}_{m \times (m+p)T_{\text{ini}}} & \mathbf{0}_{p(T_{\text{ini}}-1) \times mT_{\text{ini}}} & 
\begin{bmatrix}
\mathbf{0}_{p \times p(T_{\text{ini}}-1)} \\ 
\mathbf{I}_{p(T_{\text{ini}}-1)}
\end{bmatrix}^\top \\
\multicolumn{3}{c}{R} 
\end{array} \right],
\end{equation}
\begin{equation}
B_e = \left[\begin{array}{c}
\mathbf{0}_{m(T_{\text{ini}}-1) \times m} \\
\mathbf{I}_{m \times m} \\
\mathbf{0}_{pT_{\text{ini}} \times m} 
\end{array} \right],
\end{equation}
\end{subequations}
where $\mathbf{I}$ and $\mathbf{0}$ are identity and zero matrices with the denoted dimensions. Note that the augmented system does not require \( (A,B) \) or the exact knowledge of the system order; i.e., \( (A_e, B_e) \) are solely based on the offline dataset \eqref{eq: IO dataset}. Additionally, matrix \( R \) represents the one-step-ahead output prediction, a linear combination of sufficiently long past input-output measurements. The remaining elements in \( A_e \) and \( B_e \) are defined to shift the past input-output data for use in the next time step. In other words, this representation solely depends on past input-output measurements, the number of inputs and outputs, and the upper bound of the system’s lag.

Given the data-driven system representation matrices \( A_e \) and \( B_e \) in \eqref{eq: one-step data-driven predictor}, and the extended constraints \(\Xi\) in \eqref{eq: extended-constraints}, the maximal safe control-invariant set \(\mathcal{S}_{\Xi}\) can be computed using a set propagation technique. To compute this set, we initialize with \(\Omega_0 = \Xi\) and update it recursively at each iteration.
\begin{equation} \label{eq: iteration}
    \Omega_{k+1} = Pre(\Omega_k) \cap \Omega_k,
\end{equation}
where \( Pre(\Omega_k) = \{\xi \in \Xi \mid (\Xi \oplus (-B_e \circ \mathcal{U})) \circ A_e \} \) represents the set of all states that can transition into \(\Omega_k\) under the system dynamics and constraints. Here, \( \oplus \) denotes the Minkowski sum and \( \circ \) represents set transformation, defined as \( B_e \circ \mathcal{U} = \{B_e u \mid u \in \mathcal{U}\} \). The iterations continue until convergence, i.e., \( \Omega_{k+1} \approx \Omega_k \), at which point \( \Omega_k \) is the maximal invariant set. The MPT3 toolbox efficiently computes the maximal input-output safe set using the one-line command \texttt{sys.invariantSet()}. For further details, refer to \cite[Ch.\ 12, Ex.\ 10.6]{borrelli2017predictive}.

\section{Data-driven Input-output CBF} \label{sec: Method}

In this section, we leverage the data-driven representation \eqref{eq: one-step data-driven predictor} and the input-output safe control invariant set given in \eqref{eq: iteration} to systematically design an Input-Output Discrete-time Control Barrier Function (IO-DCBF). Inspired by discrete-time (exponential) control barrier functions in the state-space framework \cite{freire2023systematic,agrawal2017discrete}, we define the IO-DCBF as below:
\begin{definition}[IO-DCBF]\label{def: DD-CBF}  
    Let \( h: \mathbb{R}^{(p+m)T_{\text{ini}}} \to \mathbb{R} \) be a scalar continuous function defined as follows:
    \begin{equation} \label{eq: p-n CBF}
        \begin{aligned}
            h(\xi) & > 0, \quad \forall \xi \in \text{int} (\mathcal{S}), \\
            h(\xi) & = 0, \quad \forall \xi \in \partial \mathcal{S}, \\
            h(\xi) & <0, \quad \forall \xi \notin \mathcal{S}.
        \end{aligned}
    \end{equation}
The function \( h(\xi) \) is an input-output discrete-time control barrier function for system \eqref{eq: LTI-system} if, for all \( \xi_t \in \mathcal{S}\) there exists $u \in \mathcal{U}$ such that:  
    \begin{equation} \label{eq: DTECBF}  
        \Delta h(\xi, u) \geq -\lambda h(\xi),  
    \end{equation}  
    where \( \lambda \in (0, 1] \) and \( \Delta h(\xi, u) \triangleq h(A_e \xi + B_e u) - h(\xi) \). Alternatively, this condition can be expressed as:  
    \begin{equation} \label{eq: my_CBF}  
        h(A_e \xi + B_e u) \geq (1-\lambda) h(\xi).  
    \end{equation}  
\end{definition}  

Note that in discrete-time settings, \( h \) does not require to be continuously differentiable, and equation \eqref{eq: DTECBF} must hold for all \( \xi \in \mathcal{S}_\Xi \), not just for the boundary of the safe set $\mathcal{S}_\Xi$, \( \xi \in \partial \mathcal{S}_\Xi \) as in continuous-time settings. The following result follows \cite{freire2023systematic}, who showed that \( \mathcal{S} \) is invariant if and only if \( h(\xi) \) is a DCBF for $\lambda=1$. We demonstrate that the existence of an IO-DCBF leads to an input-output safe control invariant set. Conversely, the choice of $\lambda=1$ and the existence of a control invariant set imply an IO-DCBF.
\begin{proposition}
If function \( h(\xi) \) is an IO-DCBF for the system \eqref{eq: LTI-system}, then its zero-level set, \( \mathcal{S}_\Xi = \{\xi \in \Xi : h(\xi) = 0\} \), is an input-output safe control invariant set. Conversely, given an input-output safe control invariant set \( \mathcal{S}_\Xi \subseteq \Xi \) and \( \lambda = 1 \), any continuous function satisfying (\ref{eq: p-n CBF}-\ref{eq: DTECBF}) qualifies as an IO-DCBF for the system \eqref{eq: LTI-system}.
\end{proposition}
{Proof.} Let $h(\xi)$ be an IO-DCBF, then for all $\xi_t \in \mathcal{S}_\Xi$ there exists a control input $u \in \mathcal{U}$ such that $h(\xi_{t+1})\geq (1-\lambda)h(\xi_{t})\geq 0$. Therefore $\xi_{t+1} \in \mathcal{S}_\Xi$ for all $\xi_{t} \in \mathcal{S}_\Xi$, meaning that $\mathcal{S}_\Xi$, is an input-output safe control invariant set. Conversely, let $\mathcal{S}_\Xi$ to be a control invariant set then for all $\xi_t \in \mathcal{S}_\Xi$ there exists a control input $u \in \mathcal{U}$ such that $\xi_{t+1} \in \mathcal{S}$, Therefore, $h(\xi_{t+1})\geq 0$ for all $\xi_t \in \mathcal{S}_\Xi$, satisfying condition \eqref{eq: DTECBF}. This is similar to Proposition 3 in \cite{freire2023systematic}. \hfill$\square$
\begin{theorem}[Proposed IO-DCBF function] \label{th: Proposed IO-DCBF function}
Let Assumptions \ref{Assumption: I}--\ref{Assumption: II} hold, and the input-output safe control invariant set be given by \( \mathcal{S}_\Xi = \{ \xi \in \mathbb{R}^n \mid H\xi \leq c \} \). Then \( h(\xi) = \min(c - H\xi) \), where the minimum is taken element-wise over the components of the vector \( c - H\xi \), defines an IO-DCBF.
\end{theorem}
{Proof.} 
Based on Definition \ref{def: DD-CBF}, the proposed function \( h(\xi) = \min (c - H\xi) \) must satisfy the following two conditions to qualify as an IO-DCBF.

{Positivity on the zero level set:} The function \( h(\xi) = \min (c - H\xi) \) is positive for all interior points \( \xi \in \text{int}(\mathcal{S}_\Xi) \) as \( c - H\xi > \mathbf{0} \), zero for all boundary points \( \xi \in \partial\mathcal{S}_\Xi \) where \( c - H\xi = \mathbf{0} \), and negative for all points \( \xi \notin \mathcal{S}_\Xi \) since \( c - H\xi < \mathbf{0} \), where \( \mathbf{0} \) is a zero vector.

{Existence of a control input to satisfy \eqref{eq: my_CBF}:} Substituting \( h(\xi) = \min (c - H\xi) \) into \eqref{eq: my_CBF} yields  
\begin{equation} \label{eq: my_CBF_proof0}  
    \min (c - H(A_e \xi + B_e u)) \geq (1-\lambda) \min (c - H\xi).  
\end{equation}
Since \( \lambda \) belongs to \( (0,1] \) as per Definition \ref{def: DD-CBF}, setting \( \lambda = 1 \) simplifies \eqref{eq: my_CBF_proof0} to  
\begin{equation} \label{eq: my_CBF_proof1}  
    \min (c - H(A_e \xi + B_e u)) \geq 0.  
\end{equation}
The \( \min \) operator can be removed, as \( \min (c - H(A_e \xi + B_e u)) \geq 0 \) holds if and only if \( c - H(A_e \xi + B_e u) \geq \mathbf{0} \). This leads to  
\begin{equation} \label{eq: my_CBF_proof2}
    H(A_e \xi + B_e u) \leq c.
\end{equation}
By Definition \ref{def: Input-Output safe Control Invariant Set}, for all \( \xi \in \mathcal{S}_\Xi \), there always exists \( u \in \mathcal{U} \) such that \( A_e \xi + B_e u \in \mathcal{S}_\Xi \), ensuring that \eqref{eq: my_CBF_proof2} holds. Thus, the proposed function \( h(\xi) = \min (c - H\xi) \) qualifies as an IO-DCBF, satisfying \eqref{eq: DTECBF}. \hfill$\square$

In practice, as $\lambda \to 0$, the response becomes more conservative with a smoother control input, while $\lambda \to 1$ yields a less conservative response but requires a more abrupt control input. This suggests that to avoid constraint violation, for each initial condition, there exists a feasible range of $\lambda$ that can satisfy \eqref{eq: my_CBF}. Choosing $\lambda \in (0,1)$ provides an exponential lower bound for $\Delta h$, which may come at the expense of a reduced feasible set, for which the addition of a slack variable is proposed \cite{zeng2021safety,zeng2021enhancing}. In other words, any choice of $\lambda \neq 1$ may result in a compromise between feasibility and a smoother response. We define a minimally invasive IO-DCBF filter as follows in the form of Quadratic Programming (QP) with the desired decay rate $\lambda_{\text{min}}$:
\begin{subequations} \label{eq: CBF-filter for the polytopic safe set}
    \begin{align}    \underset{\lambda_{\text{min}}\leq\lambda \leq 1,u  \in \mathcal{U}}{\min} & \quad \|u - u_l\|^2 + \beta  \lambda^2  , \\
        \text{s.t.}  \quad H(A_e \xi + B_e u)& \leq c - (1 - \lambda)\min(c-H\xi) \mathbf{1},   \label{eq: Golden constraint}
    \end{align}
\end{subequations}
where $u_l$ is a potentially unsafe input, also known as a nominal or learning input. Also, $\beta$ is a large value that prioritizes the convergence rate, $\lambda$, over $\| u-u_l \|^2$. Directly imposing the barrier function \( h(\xi) = \min(c - H\xi) \) in the safety filter constraint \eqref{eq: CBF-filter for the polytopic safe set} introduces nonlinearity due to the \( \min \) operator. To obtain a QP formulation, we instead use the condition \eqref{eq: Golden constraint}, where \( \mathbf{1} \) is a constant vector with all entries equal to $1$. Note that \( \min(c-H\xi) \) is a scalar constant known at each time step as \( \xi \) is known, and the modified constraint \eqref{eq: Golden constraint} is equivalent to \eqref{eq: my_CBF}, as shown in the proof of Theorem~\ref{th: Proposed IO-DCBF function} for the case of a zero constant value. Unlike \cite{agrawal2017discrete}, the proposed safety filter \eqref{eq: CBF-filter for the polytopic safe set} ensures recursive feasibility, while, unlike \cite{freire2023systematic}, it benefits from slow braking and smoother control inputs. The proposed formulation \eqref{eq: CBF-filter for the polytopic safe set} is simpler in its implementation and interpretation than using a slack variable compared to \cite{zeng2021enhancing}.

\begin{remark}
     The input constraint \( \mathcal{U} \) is already embedded in the input-output safe set, so it does not need to be explicitly included in the safety filter formulation; the proposed CBF naturally enforces both input and output constraints.
\end{remark}

\begin{theorem} \label{Recursive feasibility} Let $\xi_{t_0} \in \mathcal{S}_\Xi$, then optimization problem \eqref{eq: CBF-filter for the polytopic safe set} is recursively feasible and the input-output trajectory satisfies input-output constraints for all $t>t_0$. 
\end{theorem}
{Proof.} 
Since $\xi_{t0} \in \mathcal{S}_\Xi$, then problem \eqref{eq: CBF-filter for the polytopic safe set} is feasible at time $t_0$ for $\lambda =1$ based on Definition \ref{def: DD-CBF}. This implies that $\xi_{t_0+1} \in \mathcal{S}_\Xi$ and $h(\xi_{t_0+1}) \geq (1-\lambda) h(\xi)$ and $\lambda_{\text{min}}\leq\lambda \leq 1$. By shifting the argument, $t_0 \rightarrow t_0+1$, it is possible to conclude that the problem \eqref{eq: CBF-filter for the polytopic safe set} stays feasible in the next time step. By relying on induction, we can conclude the problem \eqref{eq: CBF-filter for the polytopic safe set} stays feasible forever. Finally, the input-output trajectory satisfies the input-output constraints because $\xi_t \in \mathcal{S}_\Xi \subseteq \Xi$ for all $t > t_0$. \hfill$\square$
\section{Equivalence of IO-DCBF Safety Filter and Predictive Safety Filter}\label{sec: VS}
This section shows that  IO-DCBFs and MPSFs are equivalent under certain design choices and assumptions. We emphasize that this section's result is not specific to the input-output framework and can be reformulated for state-space nonlinear representations. Following is a brief introduction to MPSFs.
\subsection{Data-Driven Output Model Predictive Safety Filter}
An MPSF aims to find the nearest safe input to a potentially unsafe input by constructing a backup trajectory that reaches a terminal safe control invariant set. Specifically, it determines a control input as close as possible to the uncertified control input while ensuring the system remains safe over an infinite time horizon using a finite-time prediction horizon \cite{wabersich2018linear}. The MPSF in the input-output framework can be defined as follows \cite{bajelani2024data}:
\begin{subequations} \label{eq: IOMPSF}
\begin{align}
        &\underset{u_{[0,{N-1}]}}{\operatorname{min}}  \, \| u_0^{t}
        -u_l^{t} \|^{2}, \\
        \text{s.t.} & \quad {\xi}_{k+1}^{t} = {A_e} {\xi}_{k}^{t}+{B_e} u_{k}^{t}, \\
        & \quad {\xi}_0^{t} = {\xi}_t, \\
        & \quad \xi_k^{t} \in \Xi, \quad \forall k \in \{0, \ldots, N-1\}, \\
        & \quad u_k^{t} \in \mathcal{U}, \quad \forall k \in \{0, \ldots, N-1\}, \\
        & \quad \xi_{N}^{t} \in {\Xi_f}, \label{eq: IOMPSF-terminalset}
\end{align}
\end{subequations}
where $N$ is the prediction horizon, $\xi_k^{t}$ and $u_{k}^{t}$ are the $k^{\text{th}}$ predicted extended state and input at time $t$. The MPSF, by construction, is recursively feasible because of the terminal control invariant set, \( \Xi_f \), and can handle multiple input-output constraints. Challenges of the MPSF include quantifying uncertainty over the prediction horizon \cite{kohler2022state}, defining the terminal safe set, and managing the computational cost for long prediction horizons. In summary, the MPSF provides an implicit safe set depending on terminal constraints and the length of the prediction horizon, resulting in a more computationally expensive online solution.
\subsection{Equivalence of design in IO-DCBFs and MPSFs}
Let the terminal set in \eqref{eq: IOMPSF-terminalset} be the maximal input-output safe control invariant set \( \mathcal{S}_\Xi \), and the prediction horizon be \( N = 1 \). Also, let \( \lambda = 1 \) in \eqref{eq: CBF-filter for the polytopic safe set}. Then, the problems in \eqref{eq: IOMPSF-terminalset} and \eqref{eq: CBF-filter for the polytopic safe set} are equivalent and given by:
\begin{subequations} \label{eq: CBF-filter}
\begin{align}
    \underset{u \in \mathcal{U}}{\min} & \quad \|u - u_l\|^2, \\
    \text{s.t.} & \quad H(A_e \xi + B_e u) \leq c. \label{eq: Golden constraint2}
\end{align}
\end{subequations}
This demonstrates that the safety filter problem reduces to finding the nearest safe solution to the nominal input, \( u_l \), while ensuring the next state remains within the invariant set. Note that this result aligns with the recently introduced notion of separating hyperplanes \cite{lavanakul2024safety}[Remark 1], which can be derived from CBFs and the value function in HJB, further emphasizing the similarities between alternative formulations for safety filters. This can be seen by rewriting the constraint as \( H B_e u \leq b(\xi) \), where \( H B_e u \) and \( b(\xi) = c - H A_e \xi \) are input-dependent and extended state-dependent vectors. For a given extended state, \( H B_e u \leq b(\xi) \) defines a hyperplane, which indicates the set of safe inputs for each state.
\section{Simulation Results} \label{sec: sim}
To evaluate the proposed method, we consider a second-order LTI system with a two-step input delay and the constraints \(\mathcal{U} = \{ u \in \mathbb{R}^1 \mid |u| \leq 1 \}\), \(\mathcal{Y} = \{ y \in \mathbb{R}^1 \mid |y| \leq 1 \}\):  
\begin{equation} \label{equ: sim-ss}
    x_{t+1} = 
    \begin{bmatrix} 
        1 & 0.1 \\ 
        0 & 1 
    \end{bmatrix} x_t + 
    \begin{bmatrix} 
        0 \\ 
        0.1 
    \end{bmatrix} u_{t-2}, \quad
    y_t = 
    \begin{bmatrix} 
        1 & 0 
    \end{bmatrix} x_t.
\end{equation}
To account for an overestimation of the system's lag, \(\underline{l} = 4\), we set \(T_{\text{ini}} = 5\). It is important to note that for any \(T_{\text{ini}} \geq \underline{l}\), the input-output behavior of the system remains the same, though it results in a more computationally expensive solution. Next, an offline dataset is generated using system \eqref{equ: sim-ss}, consisting of a single input-output trajectory with 17 data points, excited by a normally distributed random input. To build the data-driven representation, the offline dataset must be reshaped into Hankel matrices $H_u \in \mathbb{R}^{6\times12}$ and $H_y \in \mathbb{R}^{6\times12}$ using \eqref{eq: Hankel matrices}. The matrix \( [H_u^{\top} H_y^{\top}]^{\top} \) should satisfy the persistent excitation (PE) assumption, with \( \text{rank}([H_u^{\top} H_y^{\top}]^{\top}) = 10 \geq T_{\text{ini}} + 1\). Therefore, the one-step ahead data-driven output prediction is defined as $y_{t}= R \xi_t$, where $R = [0.005,0.01, 0, 0, 0, 0, 0, -0.5, 0, 1.5]$. Given the input-output constraints, \( (\mathcal{U}, \mathcal{Y}) \), it is straightforward to define the extended constraints as follows for all $i=\{1,\hdots,5\}$:
\begin{equation}
    \Xi=\{\xi_t \in \mathbb{R}^{10 \times 1}| |u_{t-i}|\leq 1  , |y_{t-i}|\leq 1\}.
\end{equation}

\begin{figure}[h]
    \centering
    \includegraphics[width=0.95\linewidth]{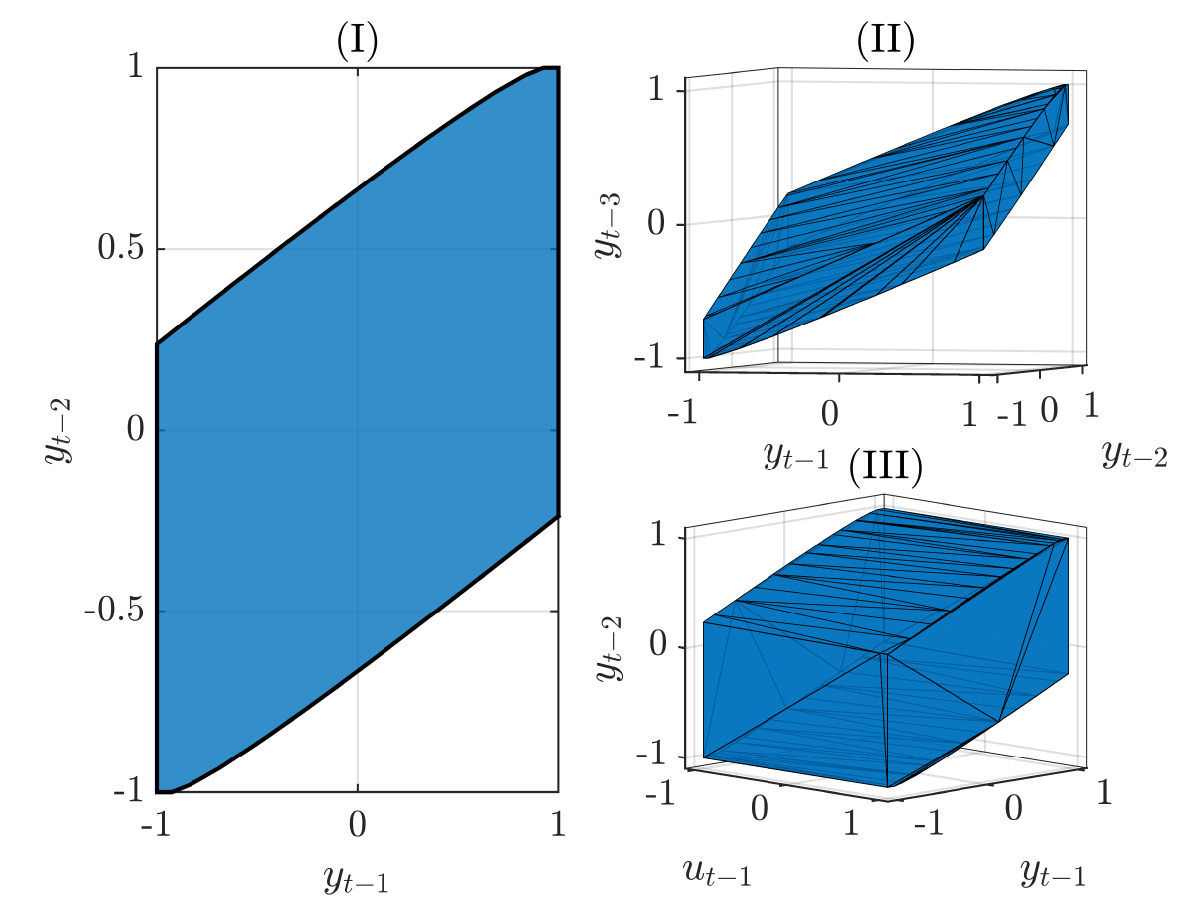} 
    \caption{The maximal input-output safe set resulting from the set propagation method: (I) projection onto \( {[y_{t-1}, y_{t-2}]}^\top \), (II) projection onto \( {[y_{t-1}, y_{t-2}, y_{t-3}]}^\top \), and (III) projection onto \( {[u_{t-1}, y_{t-1}, y_{t-2}]}^\top \).}
    \label{fig: inv-set}
\end{figure}

\begin{figure}[h]
    \centering
    \includegraphics[width=0.95\linewidth]{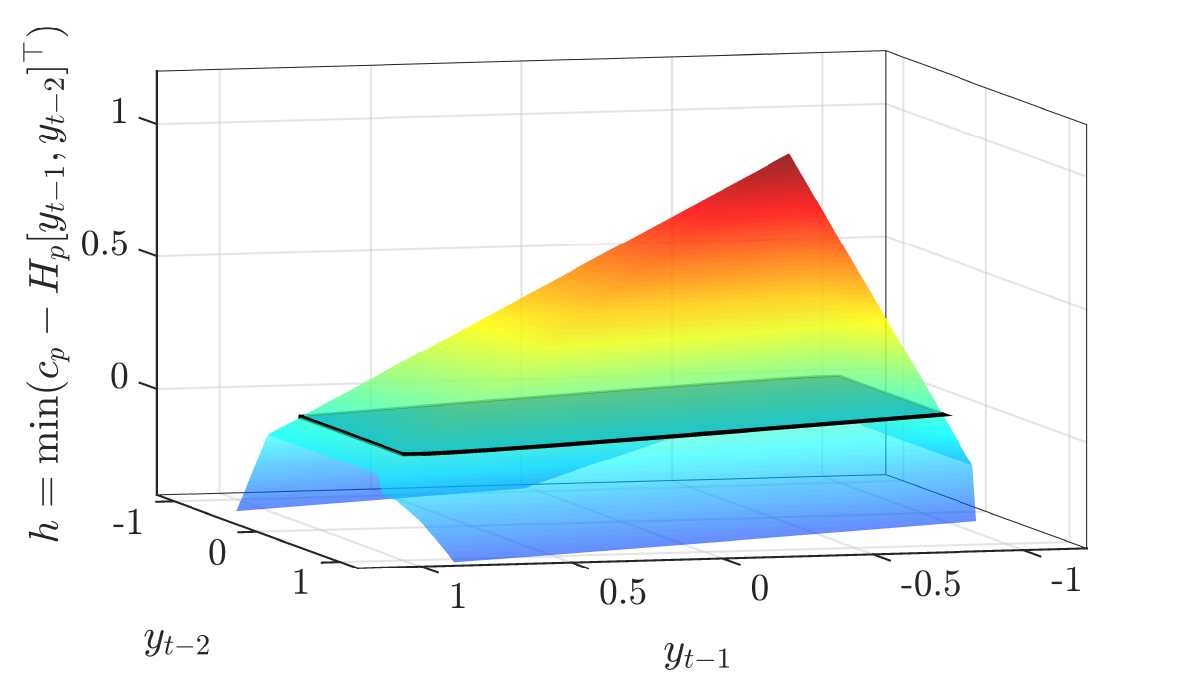} 
    \caption{The proposed control barrier function applied to the projection of the maximal input-output safe set onto \( [y_{t-1}, y_{t-2}]^\top \), where \( c_p \) and \( H_p \) define the projected set.}
    \label{fig: h_inv}
\end{figure}

We apply the standard set propagation technique to find the maximal control invariant set using the MPT3 toolbox \cite{MPT3}. The resulting set belongs to \( \mathbb{R}^{10} \). Its projections onto \( [y_{t-1}, y_{t-2}] \), \( [y_{t-1}, y_{t-2}, y_{t-3}] \), and \( [u_{t-1}, y_{t-1}, y_{t-2}] \) are displayed in Fig. \ref{fig: inv-set}. Based on the input-output safe control invariant set shown in Fig. \ref{fig: inv-set}, the proposed control barrier function is defined as \( h = \min(c - H \xi) \), where \( H \in \mathbb{R}^{120 \times 10} \) and \( c \in \mathbb{R}^{120} \) represent the H-representation of the invariant set.  Note that this representation is not unique, and the behavior of the filter, as well as its feasibility for different levels of $\lambda$, depends on the chosen representation of $h(x)$. In this example, the IO-DCBF is designed using a scaling that illustrates the recursive feasibility of the proposed filter. The IO-DCBF is shown over the projection of the invariant set onto \( [y_{t-1}, y_{t-2}] \) in Fig. \ref{fig: h_inv}. Fig. \ref{fig: h_inv} is included for illustration, and the proposed CBF depends on all five past inputs and outputs. Finally, the input-output data-driven CBF-based safety filter is defined using \eqref{eq: CBF-filter for the polytopic safe set} and evaluated for $\lambda_{\text{min}} \in \{0.01, 0.1, 1\}$ with $\beta = 10^6$.
\begin{figure}[h]
    \centering
    \includegraphics[width=0.95\linewidth]{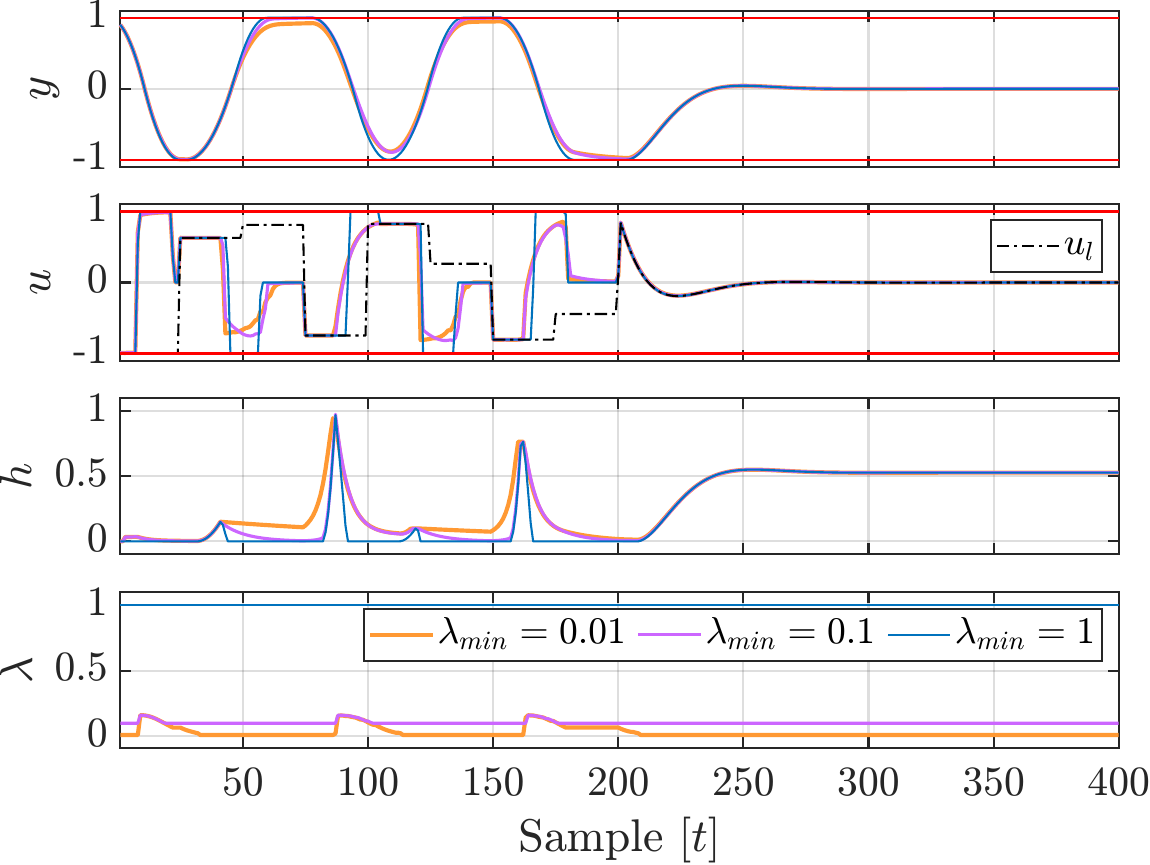} 
    \caption{Input and output of system \eqref{equ: sim-ss}, the value of the proposed barrier function, and the decay rate.}
    \label{fig: sim_results}
\end{figure}

To illustrate how the proposed safety filter minimally intervenes to keep the system safe when the nominal input is unsafe or stabilizes the origin, we assume that for \( 0 \leq t < 200 \, \text{s} \), the nominal control input \( u_l \) is a random signal with a uniform distribution and a sample time of \( 20T_s \). For \( 200 \leq t \leq 400 \, \text{s} \), the nominal control input is computed using an unconstrained LQR controller within the input-output framework (\( u^l_t = -K \xi_t \), and \( K = [0.05, 0.16, 0.15, 0.143, 0.13, 0, 0, -5.44, -5.16, 11.46] \)). The cost function is defined as $
J = \sum_{t = 0}^{\infty} \xi_t^\top\xi_t + {u^l_t}^2$ . The simulation results, including the input-output signals, the value of the proposed CBF, and variable $\lambda$ are shown in Fig. \ref{fig: sim_results}. The effect of \( \lambda_{\text{min}} \) is evident: as \( \lambda_{\text{min}} \to 1 \), the response becomes sharp, and the CBF activates near the boundaries. In contrast, as \( \lambda_{\text{min}} \to 0 \), the response smoothens, and the CBF activates farther from the boundaries. The potential infeasibility of \eqref{eq: my_CBF} for \( \lambda \neq 1 \) (e.g., \( 0.01 \) or \( 0.1 \)) is evident from the variation of \( \lambda \) illustrated in Fig.~\ref{fig: sim_results}.

\section{Discussion and Concluding Remarks}\label{sec: discussion}

This paper introduces a novel data-driven method to construct an input-output CBF, enabling safe control without reliance on explicit system models or precise knowledge of system order and delays. The proposed approach is simple yet effective, requiring only input-output measurements and supporting multiple input-output constraints without needing backup trajectories. Notably, the IO-DCBF safety filter corresponds to a model predictive safety filter with a one-step prediction horizon and a terminal set defined by the maximal safe control-invariant set. Demonstrating that the IO-DCBF safety filter is a special case of the MPSF enhances understanding of when combining MPC and CBFs is beneficial and how MPC can be improved to leverage the advantages of CBFs without adding redundant constraints or complexity.

While the method offers significant advantages, it has certain limitations. First, the assumption of an LTI system restricts its applicability to nonlinear systems. Extending this approach to nonlinear systems could involve techniques such as Koopman-based global linearization or local methods like Jacobian or feedback linearization. Compared to existing solutions, the degree of conservatism introduced by these extensions remains an open question. Second, the reliance on noise-free data poses challenges in practical scenarios. This limitation could be addressed in linear settings using zonotopic matrix identification to manage noisy data \cite{10068731}. However, effectively handling noise in linear representations of nonlinear systems is an area that requires further exploration.
 
\bibliographystyle{IEEEtran}
\bibliography{Ref}

\begin{thebibliography}{10}
\providecommand{\url}[1]{#1}
\csname url@samestyle\endcsname
\providecommand{\newblock}{\relax}
\providecommand{\bibinfo}[2]{#2}
\providecommand{\BIBentrySTDinterwordspacing}{\spaceskip=0pt\relax}
\providecommand{\BIBentryALTinterwordstretchfactor}{4}
\providecommand{\BIBentryALTinterwordspacing}{\spaceskip=\fontdimen2\font plus
\BIBentryALTinterwordstretchfactor\fontdimen3\font minus \fontdimen4\font\relax}
\providecommand{\BIBforeignlanguage}[2]{{%
\expandafter\ifx\csname l@#1\endcsname\relax
\typeout{** WARNING: IEEEtran.bst: No hyphenation pattern has been}%
\typeout{** loaded for the language `#1'. Using the pattern for}%
\typeout{** the default language instead.}%
\else
\language=\csname l@#1\endcsname
\fi
#2}}
\providecommand{\BIBdecl}{\relax}
\BIBdecl

\bibitem{breeden2023compositions}
J.~Breeden and D.~Panagou, ``Compositions of multiple control barrier functions under input constraints,'' in \emph{2023 American Control Conference (ACC)}.\hskip 1em plus 0.5em minus 0.4em\relax IEEE, 2023, pp. 3688--3695.

\bibitem{agrawal2021safe}
D.~R. Agrawal and D.~Panagou, ``Safe control synthesis via input constrained control barrier functions,'' in \emph{2021 60th IEEE Conference on Decision and Control (CDC)}.\hskip 1em plus 0.5em minus 0.4em\relax IEEE, 2021, pp. 6113--6118.

\bibitem{nguyen2016exponential}
Q.~Nguyen and K.~Sreenath, ``Exponential control barrier functions for enforcing high relative-degree safety-critical constraints,'' in \emph{2016 American Control Conference (ACC)}.\hskip 1em plus 0.5em minus 0.4em\relax IEEE, 2016, pp. 322--328.

\bibitem{ames2016control}
A.~D. Ames, X.~Xu, J.~W. Grizzle, and P.~Tabuada, ``Control barrier function based quadratic programs for safety critical systems,'' \emph{IEEE Transactions on Automatic Control}, vol.~62, no.~8, pp. 3861--3876, 2016.

\bibitem{cohen2024safety_review}
M.~H. Cohen, T.~G. Molnar, and A.~D. Ames, ``Safety-critical control for autonomous systems: Control barrier functions via reduced-order models,'' \emph{Annual Reviews in Control}, vol.~57, p. 100947, 2024.

\bibitem{clark2021verification}
A.~Clark, ``Verification and synthesis of control barrier functions,'' in \emph{2021 60th IEEE Conference on Decision and Control (CDC)}.\hskip 1em plus 0.5em minus 0.4em\relax IEEE, 2021, pp. 6105--6112.

\bibitem{prajna2007framework}
S.~Prajna, A.~Jadbabaie, and G.~J. Pappas, ``A framework for worst-case and stochastic safety verification using barrier certificates,'' \emph{IEEE Transactions on Automatic Control}, vol.~52, no.~8, pp. 1415--1428, 2007.

\bibitem{wang2024simultaneous}
X.~Wang, L.~Knoedler, F.~B. Mathiesen, and J.~Alonso-Mora, ``Simultaneous synthesis and verification of neural control barrier functions through branch-and-bound verification-in-the-loop training,'' in \emph{2024 European Control Conference (ECC)}.\hskip 1em plus 0.5em minus 0.4em\relax IEEE, 2024, pp. 571--578.

\bibitem{yaghoubi2020training}
S.~Yaghoubi, G.~Fainekos, and S.~Sankaranarayanan, ``Training neural network controllers using control barrier functions in the presence of disturbances,'' in \emph{2020 IEEE 23rd International Conference on Intelligent Transportation Systems (ITSC)}.\hskip 1em plus 0.5em minus 0.4em\relax IEEE, 2020, pp. 1--6.

\bibitem{10156243}
V.~Zinage and E.~Bakolas, ``Neural koopman control barrier functions for safety-critical control of unknown nonlinear systems,'' in \emph{2023 American Control Conference (ACC)}, 2023, pp. 3442--3447.

\bibitem{jagtap2020control}
P.~Jagtap, G.~J. Pappas, and M.~Zamani, ``Control barrier functions for unknown nonlinear systems using gaussian processes,'' in \emph{2020 59th IEEE Conference on Decision and Control (CDC)}.\hskip 1em plus 0.5em minus 0.4em\relax IEEE, 2020, pp. 3699--3704.

\bibitem{choi2021robust}
J.~J. Choi, D.~Lee, K.~Sreenath, C.~J. Tomlin, and S.~L. Herbert, ``Robust control barrier--value functions for safety-critical control,'' in \emph{2021 60th IEEE Conference on Decision and Control (CDC)}.\hskip 1em plus 0.5em minus 0.4em\relax IEEE, 2021, pp. 6814--6821.

\bibitem{freire2023systematic}
V.~Freire and M.~M. Nicotra, ``Systematic design of discrete-time control barrier functions using maximal output admissible sets,'' \emph{IEEE Control Systems Letters}, vol.~7, pp. 1891--1896, 2023.

\bibitem{cohen2024safety}
M.~H. Cohen, N.~Csomay-Shanklin, W.~D. Compton, T.~G. Molnar, and A.~D. Ames, ``Safety-critical controller synthesis with reduced-order models,'' \emph{arXiv preprint arXiv:2411.16479}, 2024.

\bibitem{molnar2023safety}
T.~G. Molnar and A.~D. Ames, ``Safety-critical control with bounded inputs via reduced order models,'' in \emph{2023 American Control Conference (ACC)}.\hskip 1em plus 0.5em minus 0.4em\relax IEEE, 2023, pp. 1414--1421.

\bibitem{folkestad2020data}
C.~Folkestad, Y.~Chen, A.~D. Ames, and J.~W. Burdick, ``Data-driven safety-critical control: Synthesizing control barrier functions with koopman operators,'' \emph{IEEE Control Systems Letters}, vol.~5, no.~6, pp. 2012--2017, 2020.

\bibitem{9654975}
F.~Fiedler and S.~Lucia, ``On the relationship between data-enabled predictive control and subspace predictive control,'' in \emph{2021 European Control Conference}, 2021, pp. 222--229.

\bibitem{berberich2020robust}
J.~Berberich, J.~K{\"o}hler, M.~A. M{\"u}ller, and F.~Allg{\"o}wer, ``Robust constraint satisfaction in data-driven {MPC},'' in \emph{2020 59th IEEE Conference on Decision and Control (CDC)}.\hskip 1em plus 0.5em minus 0.4em\relax IEEE, 2020, pp. 1260--1267.

\bibitem{willems2005note}
J.~C. Willems, P.~Rapisarda, I.~Markovsky, and B.~L. De~Moor, ``A note on persistency of excitation,'' \emph{Systems \& Control Letters}, vol.~54, no.~4, pp. 325--329, 2005.

\bibitem{markovsky2008data}
I.~Markovsky and P.~Rapisarda, ``Data-driven simulation and control,'' \emph{International Journal of Control}, vol.~81, no.~12, pp. 1946--1959, 2008.

\bibitem{berberich2021design}
J.~Berberich, J.~K{\"o}hler, M.~A. M{\"u}ller, and F.~Allg{\"o}wer, ``On the design of terminal ingredients for data-driven {MPC},'' \emph{IFAC-PapersOnLine}, vol.~54, no.~6, pp. 257--263, 2021.

\bibitem{pmlr-v242-bajelani24a}
M.~Bajelani and K.~V. Heusden, ``From raw data to safety: {R}educing conservatism by set expansion,'' in \emph{Proceedings of the 6th Annual Learning for Dynamics \& Control Conference}, ser. Proceedings of Machine Learning Research, vol. 242.\hskip 1em plus 0.5em minus 0.4em\relax PMLR, 15--17 Jul 2024, pp. 1305--1317.

\bibitem{qin2024data}
Z.~Qin and A.~Karimi, ``Data-driven output feedback control based on behavioral approach,'' in \emph{2024 American Control Conference (ACC)}.\hskip 1em plus 0.5em minus 0.4em\relax IEEE, 2024, pp. 3954--3959.

\bibitem{borrelli2017predictive}
F.~Borrelli, A.~Bemporad, and M.~Morari, \emph{Predictive control for linear and hybrid systems}.\hskip 1em plus 0.5em minus 0.4em\relax Cambridge University Press, 2017.

\bibitem{agrawal2017discrete}
A.~Agrawal and K.~Sreenath, ``Discrete control barrier functions for safety-critical control of discrete systems with application to bipedal robot navigation.'' in \emph{Robotics: Science and Systems}, vol.~13.\hskip 1em plus 0.5em minus 0.4em\relax Cambridge, MA, USA, 2017, pp. 1--10.

\bibitem{zeng2021safety}
J.~Zeng, B.~Zhang, Z.~Li, and K.~Sreenath, ``Safety-critical control using optimal-decay control barrier function with guaranteed point-wise feasibility,'' in \emph{2021 American Control Conference (ACC)}.\hskip 1em plus 0.5em minus 0.4em\relax IEEE, 2021, pp. 3856--3863.

\bibitem{zeng2021enhancing}
J.~Zeng, Z.~Li, and K.~Sreenath, ``Enhancing feasibility and safety of nonlinear model predictive control with discrete-time control barrier functions,'' in \emph{2021 60th IEEE Conference on Decision and Control (CDC)}.\hskip 1em plus 0.5em minus 0.4em\relax IEEE, 2021, pp. 6137--6144.

\bibitem{wabersich2018linear}
K.~P. Wabersich and M.~N. Zeilinger, ``Linear model predictive safety certification for learning-based control,'' in \emph{2018 IEEE Conference on Decision and Control (CDC)}.\hskip 1em plus 0.5em minus 0.4em\relax IEEE, 2018, pp. 7130--7135.

\bibitem{bajelani2024data}
M.~Bajelani and K.~van Heusden, ``Data-driven safety filter: An input-output perspective,'' in \emph{2024 American Control Conference (ACC)}.\hskip 1em plus 0.5em minus 0.4em\relax IEEE, 2024, pp. 5106--5112.

\bibitem{kohler2022state}
J.~K{\"o}hler, K.~P. Wabersich, J.~Berberich, and M.~N. Zeilinger, ``State space models vs. multi-step predictors in predictive control: Are state space models complicating safe data-driven designs?'' in \emph{2022 IEEE 61st Conference on Decision and Control (CDC)}.\hskip 1em plus 0.5em minus 0.4em\relax IEEE, 2022, pp. 491--498.

\bibitem{lavanakul2024safety}
W.~Lavanakul, J.~Choi, K.~Sreenath, and C.~Tomlin, ``Safety filters for black-box dynamical systems by learning discriminating hyperplanes,'' in \emph{6th Annual Learning for Dynamics \& Control Conference}.\hskip 1em plus 0.5em minus 0.4em\relax PMLR, 2024, pp. 1278--1291.

\bibitem{MPT3}
M.~Herceg, M.~Kvasnica, C.~Jones, and M.~Morari, ``{Multi-Parametric Toolbox 3.0},'' in \emph{Proc.~of the European Control Conference}, Z\"urich, Switzerland, July 17--19 2013, pp. 502--510.

\bibitem{10068731}
A.~Alanwar, A.~Koch, F.~Allgöwer, and K.~H. Johansson, ``Data-driven reachability analysis from noisy data,'' \emph{IEEE Transactions on Automatic Control}, vol.~68, no.~5, pp. 3054--3069, 2023.

\end{thebibliography}

\end{document}